\documentclass[a4paper,11pt]{article}

\pdfoutput=1

\usepackage{jheppub}
\usepackage{amsfonts}
\usepackage{amsmath}
\usepackage{amssymb}
\usepackage{graphicx,color}
\usepackage{float}
\usepackage{hyperref}
\usepackage[dvipsnames]{xcolor}
\usepackage{caption}
\usepackage{subcaption}

\usepackage{soul}

\preprint{MIT-CTP/5905}

\newcommand{\Mp}{M_{\rm Pl}} %Planck mass%

\usepackage{xcolor}
\setcitestyle{square, numbers, comma, sort&compress}

\title{\boldmath Heterotic Warm Inflation }

\author[a]{Arjun Berera}    
\author[b]{, Heliudson Bernardo}
\author[a,c]{, Suddhasattwa Brahma}
\author[d]{, Jaime Calder\'on-Figueroa}
\author[e]{, Rudnei O. Ramos}    
\author[f]{and Michael W. Toomey}

\affiliation[a]{Higgs Centre for Theoretical Physics, School of Physics and Astronomy, University of Edinburgh, Edinburgh EH9 3FD, UK}
\affiliation[b]{Department of Physics \& Astronomy and Quantum Horizons Alberta, University of Lethbridge, \protect\\
4401 University Drive, Lethbridge AB, T1K 3M4, Canada}
\affiliation[c]{Physics and Applied Mathematics Unit, Indian Statistical Institute, 203 B.T. Road, Kolkata 700108, India}
\affiliation[d]{Astronomy Centre, University of Sussex, Falmer, Brighton, BN1 9QH, UK}
\affiliation[e]{Departamento de F\'{\i}sica Te\'orica,
  Universidade do Estado do Rio de Janeiro,
  20550-013 Rio de Janeiro, RJ, Brazil}
\affiliation[f]{Center for Theoretical Physics -- a Leinweber Institute, Massachusetts Institute of Technology, Cambridge, MA 02139, USA}

\emailAdd{ab@ed.ac.uk}
\emailAdd{heliudson.bernardo@uleth.ca}
\emailAdd{sbrahma@exseed.ed.ac.uk}
\emailAdd{jrc43@sussex.ac.uk}
\emailAdd{rudnei@uerj.br}
\emailAdd{mtoomey@mit.edu}

\abstract{ 
We propose a two-field model of warm inflation motivated by a heterotic string construction, 
involving an axion and a dilaton-like scalar field with non-trivial kinetic mixing. 
Gauge-field interactions generate dissipation and thermal corrections affecting both fields.
A systematic numerical analysis reveals a range of dynamical regimes, including effectively 
single-field and multi-field behavior. We find that warm inflation is typically realized 
along the axion direction, while thermal corrections tend to hinder sustained dilaton-driven 
inflation over most of the parameter space. Although configurations exist in which the 
dilaton becomes dynamically relevant, particularly near the end of inflation, the majority 
of viable solutions are effectively single-field and axion-dominated.
These results point to a dynamical mechanism in heterotic-inspired models that naturally 
favors axion-driven warm inflation while limiting the role of the dilaton.
} 
\begin{document} 
\maketitle

%\flushbottom

%%%%%%%%%%%%%%%%%%%%%%%%%%%%%%%%%%%%%%%%  
\section{Introduction}

Cosmic inflation~\cite{Kazanas:1980tx, Guth:1980zm, Sato:1981ds, Sato:1980yn, Linde:1981mu, Albrecht:1982wi}, a period of exponential expansion in the universe's earliest moments, resolves fundamental puzzles left unanswered by the standard Big Bang model. This explains the striking large-scale uniformity of the universe (the horizon problem), its nearly flat geometry (the flatness problem), and the absence of exotic relics (the monopole problem). The stretching of quantum fluctuations during the inflationary phase leads to a mechanism for which cosmic seeds for galaxies and large-scale structures can be formed, while also making possible the imprinting of detectable patterns in the cosmic microwave background (CMB). By bridging quantum mechanics and cosmology, inflation not only harmonizes observed features of the cosmos, but also provides a framework for testing the primordial origins of the universe through precision measurements of CMB radiation and large-scale structure ~\cite{Planck:2018jri, Planck:2019nip, ACT:2025fju, Euclid:2023shr}.

However, inflationary model-building consists of not only crafting a flat enough potential for the inflaton to slowly roll down, but also protecting its flatness against quantum corrections. Moreover, resorting to a fundamental theory is essential for explaining the origins of the inflaton field itself. Constructing inflation models within string theory offers a compelling pathway towards finding elegant solutions for both these problems. It gives a consistent UV-completion for early universe cosmology, addressing fundamental questions left open by effective field theory (EFT) approaches (see, e.g. \cite{Baumann:2014nda}, or \cite{Cicoli:2023opf, Brandenberger:2023ver} and references therein for a recent review). The rich geometric landscape of string theory, with its vast array of compactifications, branes, and fluxes, naturally gives rise to scalar fields with specific symmetries (such as axions) that can drive inflationary dynamics while remaining consistent with quantum gravity constraints, such as the absence of trans-Planckian field excursions~\cite{Grana:2005jc, Blumenhagen:2006ci, Svrcek:2006yi}. By embedding inflation into string theory, we gain a framework to explore observable signatures tied to higher-dimensional geometry or quantum gravity effects, while testing the viability of string theory through cosmological observations such as CMB measurements~\cite{Planck:2018jri}.

Warm inflation (WI)~\cite{Berera:1995wh, Berera:1995ie,Berera:1996nv,Berera:1996fm, Berera:1998px} presents a dynamic alternative to conventional cold inflation by incorporating continuous energy exchange between the inflaton field and a bath of particles during the accelerated expansion phase (for reviews, see~\cite{Berera:2008ar, BasteroGil:2009ec, Kamali:2023lzq, Berera:2023liv}). 
The name of the dynamics emphasizes the key distinguishing thermodynamic property that the state of the universe is warm through possession of a radiation bath of particles. Unlike traditional models, where inflation ends abruptly and reheating must generate primordial plasma, WI naturally dissipates the energy of the inflaton into radiation through particle interactions, with models studied up to now sustaining a near-thermal environment throughout. This framework addresses key challenges, such as alleviating the need for fine-tuned potentials or ad hoc prescriptions to keep the inflaton potential's curvature small (the so-called $\eta$-problem), as well as eliminating the need to postulate some `reheating' mechanism, while offering distinct observational signatures---such as enhanced curvature perturbations or suppressed tensor 
modes---linked to dissipative dynamics~\cite{Bartrum:2013fia, Benetti:2016jhf, Bastero-Gil:2016qru, Bastero-Gil:2014raa, Mirbabayi:2022cbt}. Comparison to CMB data has shown that WI provides a good fit, as illustrated in some recent analysis in \cite{Montefalcone:2022jfw,Kumar:2024hju,Santos:2024plu, Berera:2025vsu, ORamos:2025uqs}. By embedding inflation within a near thermalized environment, WI bridges early-universe physics with realistic particle physics models, providing testable predictions for CMB anomalies or small-scale structure, and revitalizing the search for a unified, thermodynamically consistent origin of cosmic structure. Note that the near thermalization feature is not inherent in the description of WI, and systems in much further non-equilibrium  states in principle could also be possible, but the near thermal limit in practical terms is the most amenable to calculation and the one primarily studied to date for WI (for some related work, see~\cite{Dymnikova:2000gnk,Dymnikova:2001jy,Kamada:2009hy,Levy:2020zfo}).

In this paper, we present a WI model based on an heterotic string construction \cite{Gross:1984dd, Green:1984sg, Hull:1985jv} (for earlier WI models motivated by string theory, see, e.g. refs~\cite{Berera:1998px, Berera:1999wt, Motaharfar:2021egj, Chakraborty:2025yms}). The model has many attractive features. It describes a two-field WI realization where dissipation terms can be defined consistently through the existing interactions of the axion-like and dilaton scalar fields with a non-Abelian gauge field. Both fields and interactions emerge naturally from the model construction. One of the main novelty of this model lies in a kinetic coupling between the dilaton and the axion fields, which is a generic feature of working with complex axio-dilaton moduli in string compactifications \cite{Candelas:1985en, Strominger:1986uh, Dasgupta:1999ss, Giddings:2001yu}. This feature originates from the no-scale nature of the resulting K\"ahler potential in four dimensions after dimensional reduction \cite{Witten:1985xb}. Although such kinetic coupling has always been present in low-energy, four-dimensional actions in string theory, its usage in phenomenological models has increased recently \cite{Bernardo:2022ztc, Alexander:2022own, Gallego:2024gay, Smith:2024ibv, Ortega:2024prv, Toomey:2025mvx}. We identify regimes in which the model can behave simply as `minimal WI'~\cite{Berghaus:2019whh}, where the inflation proceeds along the axion field direction and the dilaton energy density remains much smaller than the radiation energy density throughout the dynamics. However, there can also be parameter regimes in which the background expansion is essentially driven by cold inflation. This shows how dissipative dynamics within a realistic model can be determined depending on the strength of the interactions between the different fields.

This paper is organized as follows. In section~\ref{section2}, we give a preliminary  setup of the model studied here, with emphasis on how the kinetic coupling changes the dynamics of WI. In section~\ref{section3}, we review the heterotic string construction motivating the WI model presented here. In section~\ref{section4}, we present the WI model as motivated by the heterotic string construction. The dissipation terms for the axion-like and dilaton fields are explicitly derived. The thermal contributions resulting from the interactions of the scalar fields with the nonabelian gauge field are also derived and their effects on the dynamics are made explicit. In section~\ref{section5}, we present the numerical results for the combined background dynamics of the two scalar fields with the radiation bath. Comments about the expected perturbations for the model are briefly discussed in section~\ref{sec6}. Our conclusions are presented in section~\ref{conclusions}. We use the mostly plus metric signature and adopt Planckian units unless otherwise stated.

%%%%%%%%%%%%%%%%%%%%%%%%%%%%%%%%%%%%%%%%  
\section{Motivation}
\label{section2}

Motivated by the general axion-saxion kinetic coupling in four-dimensional supergravity models derived from string theory, we wish to study how a kinetic mixing between two scalar fields affects the dynamics of WI. 

The class of models in which we are interested has the form
\begin{align}\label{Theaction}
    S =& \int d^4 x \sqrt{-g}\left[\frac{1}{2\kappa^2} R - \frac{1}{2}g^{\mu\nu}\partial_\mu \chi \partial_\nu \chi -f(\chi)\frac{1}{2}g^{\mu\nu} \partial_\mu\phi \partial_\nu \phi - V(\chi, \phi)  \right]\nonumber\\
    &+ S_{\text{thermal bath}} + S_{\text{coupling}}\,,
\end{align}
where $\kappa\equiv 1/M_{\rm Pl}$, with $M_{\rm Pl} \simeq 2.4 \times 10^{18}$ GeV is the reduced Planck mass. The first line of eq.~(\ref{Theaction}) was used in \cite{Alexander:2022own} to describe the early dark energy and address the $H_0$ tension. 
As we shall see in the next section, string theory supports interpreting $\phi$ as an axion and $f(\chi) = e^{-\lambda \kappa \chi}$ with $|\lambda| \sim \mathcal{O}(1)$.

The second line in the action \eqref{Theaction} represents possible kinetic terms for the degrees of freedom corresponding to the thermal bath of WI and the coupling of $\chi$ and $\phi$ to it. Without specifying the second line of \eqref{Theaction}, but assuming thermalization, we can get the equations of motion for our system from local conservation of its energy and momentum,
\begin{equation}\label{conservationeq}
    \nabla^\mu \left( T_{\mu\nu} + T^{\text{tb}}_{\mu\nu}\right) = 0,
\end{equation}
where $T_{\mu\nu}$ is the energy-momentum as computed from the action for $\chi$ and $\phi$ (but possibly written in terms of ``renormalized'' fields and potential, due to thermalization),
\begin{equation}
    T_{\mu\nu} = \partial_\mu \chi \partial_\nu \chi + f(\chi) \partial_\mu \phi \partial_\nu \phi - g_{\mu\nu} \left[\frac{1}{2}(\partial \chi)^2 + \frac{1}{2}f(\chi)(\partial \phi)^2 + V\right]\,, 
\end{equation}
while $T^{\text{tb}}_{\mu\nu}$ is the energy-momentum tensor of the thermal bath, which we assume can be described by a perfect fluid. Due to the coupling between the $(\chi,\phi)$ system and the thermal bath, we have
\begin{equation}
    \nabla_\mu T^{\mu\nu}_{\text{tb}} = J^\nu_{(\phi)} + J^\nu_{(\chi)}\,,
\end{equation}
where the vectors $J^\nu_{(\phi)}$ and $J^\nu_{(\chi)}$ describe the energy-fluxes from the fields $\phi$ and $\chi$ to thermal bath, respectively. Then, the conservation equation \eqref{conservationeq} gives
\begin{align}
    0=\nabla^\mu \left(T_{\mu\nu} + T^{\text{tb}}_{\mu\nu} \right) &=  \Box \chi \partial_\nu \chi +2\partial^\mu \chi [\nabla_\mu, \nabla_\nu] \chi + f' \partial^\mu \chi \partial_\mu \phi \partial_\nu \phi + f\Box \phi \partial_\nu \phi + \nonumber\\
    &+2 f\partial^\mu \phi [\nabla_\mu, \nabla_\nu] \phi - \frac{1}{2}f' (\partial \phi)^2 \partial_\nu \chi - V_\chi \partial_\nu \chi - V_\phi \partial_\nu \phi +J^{(\phi)}_\nu + J^{(\chi)}_\nu\,.
\end{align}
Using the fact that $[\nabla_\mu,\nabla_\nu] \phi = 0 = [\nabla_\mu,\nabla_\nu] \chi$ (in the absence of torsion and for continuous scalar configurations), contracting the above equation with $\partial^\nu \chi$, and simplifying, we find
\begin{equation}
    (\partial \chi)^2 \left[\Box \chi - \frac{1}{2}f' (\partial \phi)^2 - V_\chi \right] + (\partial^\nu \chi \partial_\nu \phi)\left[f\Box \phi + f'(\partial^\mu \chi \partial_\mu \phi)- V_\phi \right] +\partial^\nu \chi J_{\nu}^{(\chi)} + \partial^\nu \chi J_{\nu}^{(\phi)} = 0\,. 
\end{equation}
Now, assuming $J_\nu^{(\phi)} = \Theta_\phi \partial_\nu \phi$ and $J_\nu^{(\chi)} = \Theta_\chi \partial_\nu \chi$, where $\Theta_i$ can be thought of as quantifying the energy transfer for the two fields, we have
\begin{equation}
    (\partial \chi)^2 \left[\Box \chi - \frac{1}{2}f' (\partial \phi)^2 - V_\chi + \Theta_\chi \right] + (\partial^\nu \chi \partial_\nu \phi)\left[f\Box \phi + f'(\partial^\mu \chi \partial_\mu \phi)- V_\phi + \Theta_\phi \right]  = 0\,.
\end{equation}
However, this can only hold for any field configuration provided that
\begin{subequations}
    \begin{align}
        \Box \chi - \frac{1}{2}f' (\partial \phi)^2 - V_\chi + \Theta_\chi &= 0\,, \\
        f\Box \phi + f'(\partial^\mu \chi \partial_\mu \phi)- V_\phi + \Theta_\phi &=0\,,
    \end{align}
\end{subequations}
to which we should also include the conservation equation,
\begin{equation}
    \nabla^\mu T_{\mu\nu}^{\text{tb}} = \Theta_\phi \partial_\nu \phi + \Theta_\chi \partial_\nu \chi\,.
\end{equation}
Assuming a perfect fluid form for $T_{\mu\nu}^{\text{tb}}$ and contracting the above with the fluid's velocity $U^\mu$ yields the following
\begin{equation}
    U^\mu \nabla_\mu \rho + (\rho +p)\nabla_\mu U^\mu = - \Theta_\chi U^\mu \partial_\mu \chi - \Theta_\phi U^\mu \partial_\mu \phi\,.
\end{equation}

For a flat FLRW background, assuming homogeneous fields and going to the thermal-bath rest frame, we have that
\begin{subequations}
    \begin{align}
        \Ddot{\chi} + 3H \dot{\chi} - \frac{1}{2}f' \dot{\phi}^2 + V_\chi - \Theta_\chi &=0\; ,\\
        f\left(\Ddot{\phi} + 3H\dot{\phi}\right) + f' \dot{\phi} \dot{\chi} + V_\phi - \Theta_\phi &=0\;,\\
        \dot{\rho} +3(\rho +p)H &= -\Theta_\chi \dot{\chi} - \Theta_\phi \dot{\phi}   \;,
    \end{align}
\end{subequations}
where $H$ is the Hubble rate of expansion and the dot denotes the derivative with respect to cosmic time. If we further assume that the energy transfer functions $\Theta_i$ are proportional to the field's velocities, $\Theta_\phi = - \Upsilon_\phi \dot{\phi}$ and $\Theta_\chi = - \Upsilon_\chi \dot{\chi}$, we finally have
\begin{subequations}
\label{eq:KMIX_WI_model_eom}
     \begin{align}
        \Ddot{\chi} + 3H \dot{\chi} - \frac{1}{2}f' \dot{\phi}^2 + V_\chi &= - \Upsilon_\chi \dot{\chi}\,,\\
        f\left(\Ddot{\phi} + 3H\dot{\phi}\right) + f' \dot{\phi} \dot{\chi} + V_\phi &=- \Upsilon_\phi \dot{\phi}\,,\\
        \dot{\rho} +3(\rho +p)H &= \Upsilon_\chi \dot{\chi}^2+ \Upsilon_\phi \dot{\phi}^2   \,.
    \end{align}   
\end{subequations}
The system \eqref{eq:KMIX_WI_model_eom} is a multifield one, with a kinetic coupling between the fields, a situation not considered in the minimal WI scenario.
For a thermal radiation bath, this set of equations was considered in \cite{Sa:2020fvn} (with a certain choice of $V$ and exponential form for $f(\chi)$; more importantly, the dissipation coefficients were put in by hand). From the equations of motion \eqref{eq:KMIX_WI_model_eom}, we see that the kinetic coupling makes $\dot{\phi}^2$ act as a source for $\chi$, and $\dot{\chi}$ appears as a ``friction'' term\footnote{The interpretation of the $f'\dot{\chi}\dot\phi$-term as a friction term in the equation of motion for $\phi$ requires $\dot{f}>0$. More importantly is the direction of the energy flow between $\phi$ and $\chi$, which is determined by $\text{sign}(f'(\chi))$. As shown in \cite{Alexander:2022own}, the energy flows from $\phi$ to $\chi$ asymptotically in field space and regardless of $\text{sign}(f'(\chi))$.} in the equation of motion of $\phi$. Moreover, both fields are coupled to the thermal bath, such that they source $\rho$ while dissipating energy via the non-vanishing $\Upsilon_{\phi, \chi}$ coefficients. Previous similar cases of multifield models include, for example \cite{Braglia:2020eai, Langlois:2008mn, Karamitsos:2017elm, DiMarco:2002eb} for kinetically mixed models in the context of cosmology (but without a thermal bath), \cite{Li:2019zbk, Wang:2018cev} for multifield WI (but without kinetic coupling), and \cite{Brinkmann:2022oxy} for multifield quintessence models with kinetic mixing motivated by string theory.
See also \cite{Kamali:2019ppi} for an earlier study where WI is driven by a pseudoscalar inflaton with a gauge field coupling, somewhat akin to our axion direction, though without the specific heterotic context and without a spectator dilaton field.

%%%%%%%%%%%%%%%%%%%%%%%%%%%%%%%%%%%%%%%%  
\section{Heterotic string origin of the model}
\label{section3}

In this section, we will explain how to obtain an action of the form \eqref{Theaction} from heterotic string theory along with the following terms for the thermal bath and the coupling between the scalars and the bath of the form:
\begin{subequations}
\begin{align}
    S_{\text{thermal bath}} &= -\frac{1}{2g^2} \int \text{tr}\; F\wedge *F \,, \\
    S_{\text{coupling}} &= \beta \int e^\chi\; \text{tr}\; F\wedge *F + \beta \int \phi \; \text{tr}\; F\wedge F \,,
\end{align}
\end{subequations}
where $F$ is the field strength of some gauge fields  and $g$ is the associated coupling parameter.

We start with the action for the massless bosonic spectrum of heterotic string theory at weak coupling \cite{Polchinski:1998rr} (in the Einstein frame),
\begin{equation}\label{heteroticaction}
    S = \frac{1}{2\kappa^2_{10}}\int d^{10}x \sqrt{-G}\left[R - \frac{1}{2}(\partial \varphi)^2 - \frac{e^{-\varphi}}{2}|\Tilde{H}_3|^2 - \frac{\kappa^2_{10}}{30g^2_{10}}e^{-\varphi/2}\text{tr}|F_2|^2\right]\,, 
\end{equation}
with 
\begin{equation}
    dF_2 = 0 , \quad \Tilde{H}_3 = H_3 - \frac{\kappa^2_{10}}{g^2_{10}}\left(\Omega_3(A)- \Omega_3(w)\right)\,,
\end{equation}
where the $\Omega_3$ terms are the Chern-Simons terms for the gauge-field $A_1 = A_\mu dx^\mu$ and spin connection $\omega_1 = \omega_\mu dx^\mu$:
\begin{equation}
    \Omega_3(A) = \frac{1}{30}\text{tr}\left(A_1 \wedge d A_1 -i \frac{2}{3}A_1\wedge A_1 \wedge A_1\right)\,, \;\; \Omega_3(\omega) = \text{tr}\left(\omega_1 \wedge d \omega_1 + \frac{2}{3} \omega_1 \wedge \omega_1 \wedge \omega_1\right)\,.
\end{equation}
In the action \eqref{heteroticaction} above, $\varphi$ is the dilaton, $H_3=dB_2$ is the field strength for a 2-form field $B_2$, and $F_2=dA_1-i A_1 \wedge A_1$ is the field strength for a non-abelian gauge field $A_1$ in the adjoint of an $\text{E}_8 \times \text{E}_8$ or SO$(32)$ gauge group (which justify the factors of $1/30$ accompanying some traces in this Section.) . The trace acting on the gauge fields is with respect to the adjoint representation, while the trace acting on the spin connection is in the vector representation of $SO(1,9)$. We also have $2\kappa_{10}^2= (2\pi)^7 \alpha'^4$ and $\kappa_{10}^2/g_{10}^2 = \alpha'/4$, where $\sqrt{\alpha'}$ is the string length that defines the string scale $M_s = \alpha'^{-1/2}$. Note that $B_2$ and $\varphi$ are dimensionless.

The Bianchi identity for $\Tilde{H}_3$ is
\begin{equation}\label{Bianchiid}
    d\Tilde{H}_3 = \frac{\kappa^2_{10}}{g^2_{10}}\left(\frac{1}{30}\text{tr}F_2\wedge F_2 - \text{tr}R_2 \wedge R_2\right)
\end{equation}
where we are using the following notation for forms,
\begin{equation}
    \int F_p \wedge * F_p = \int d^D x \sqrt{-G} |F_p|^2, \quad |F_p|^2 =\frac{1}{p!}G^{a_1 b_1}\cdots G^{a_p b_p} F_{a_1 \cdots a_p}F_{b_1 \cdots b_p}\,,
\end{equation}
and the components of the Hodge dual of a $p$-form are
\begin{equation}
    (*A_p)_{a_1 \cdots a_{D-p}} = \frac{1}{p!}\epsilon_{a_1\cdots a_{D-p}}^{\;\;\;\;\;\;\;\;\;\;\;\;b_1\cdots b_p}A_{b_1 \cdots b_p}\,.
\end{equation}

The shift in $H_3$ by the Chern-Simons forms is a consequence of the Green-Schwarz mechanism for ten-dimensional anomaly cancellation~\cite{Green:1984sg,Green:2012pqa}. This also requires the following extra terms in the action~\cite{Polchinski:1998rr}
\begin{equation}\label{anomalyterm}
    S \supset -\frac{1}{768}\int B \wedge \left[\text{tr} R^4 +\frac{1}{4} \text{tr} R^2 \text{tr} R^2 -\frac{1}{30} \text{tr}F^2\text{tr}R^2 +\frac{1}{3}\text{tr}F^4 -\frac{1}{900}\text{tr}F^2 \text{tr}F^2\right] \;, 
\end{equation}
where the powers in the curvature two-forms denote wedge products, e.g. $F^n = F\wedge\dots \wedge F$. We shall see that these one-loop anomaly-induced terms are crucial to getting the coupling with axions and gauge fields in the lower-dimensional EFT.

%%%%%%%%%%%%%%%%%%%%%%%%%%%%%%%%%%%%%%%%  
\subsection{Heuristics of compactification}

To make this work self-contained and broadly accessible, in this section we perform the dimensional reduction of the ten-dimensional heterotic string theory action \eqref{heteroticaction} on a class of simplified internal manifolds. This approach allows us to understand the higher-dimensional origin of the terms in the four-dimensional action which are relevant for our model, without involving unnecessary complications. However, our complete model is defined in the next section by employing a more systematic way to obtain the four-dimensional low-energy action from \eqref{heteroticaction} (see, e.g., \cite{Cicoli:2013rwa} for a review on heterotic string compactification).

To obtain an EFT in four dimensions, we assume that the spacetime is a product of four- and six-dimensional manifolds $M_{10} = M_4 \times M_6$, the latter being a compact one. We then focus on the massless fields in four dimensions (Kaluza-Klein truncation), which correspond to the zero modes of the internal manifold. We expect the following massless scalar spectrum in four dimensions: a four-dimensional dilaton, a scalar dual to $\Tilde{H}_{\rho\mu\nu}$, scalars from $B_{mn}$ and $A_m$, and scalars corresponding to the size and shape deformation of the internal space. To give a taste of the dimensional reduction procedure, consider reducing with the metric ansatz
\begin{equation}
    ds^2 = G_{ab}(x^c)dx^a dx^b = e^{-6\sigma(x)}g_{\mu\nu}(x^\rho)dx^\mu dx^\nu + e^{2\sigma(x)}h_{mn}(y)dy^m dy^n\;,
\end{equation}
where the factors of $\sigma(x)$ are necessary to ensure the correct normalization of the four-dimensional Einstein-Hilbert action (corresponding the 4-d metric, $g_{\mu\nu}$). With the above ansatz, we assume that only one modulus will be associated with the internal manifold -- its overall size, corresponding to $\sigma$. Moreover, since $B_{mn}(x)$ and $A_{m}(x)$ configurations should also satisfy the ten-dimensional equations of motion, we will get four-dimensional massless scalars provided they correspond to harmonic forms in the internal space. The number of harmonic $p$-forms admitted in the internal space is its Betti number $b_p(M_6)$. We shall neglect the $A_m$ moduli and assume $b_{2}(M_6)=1$, such that there is a modulus associated with $B_{mn}$ with $m$ and $n$ taking values in the 2-cycle direction, say $m=4$ and $n=5$. With all these assumptions, we are interested in the four-dimensional theory for four scalar fields: the dilaton, the dual to $\tilde{H}_{\rho\mu\nu}$, the size of the internal space, and $B_{45}$.

After straightforward computations, the dimensional reduction of the gravity-scalar part of \eqref{Theaction} gives
\begin{equation}
    S = \frac{V_6}{2\kappa^2_{10}} \int d^4x \sqrt{-g}\left[R(g) + e^{-4\sigma}\langle R(h)\rangle -24\partial_\mu \sigma \partial^\mu \sigma -\frac{1}{2}\partial_\mu \varphi \partial^\mu \varphi+\cdots \right]\,,
\end{equation}
where $V_6$ is the fiducial internal volume and $\langle R(h)\rangle$ is the mean curvature of the internal space:
\begin{equation}
    \langle R(h)\rangle = \frac{1}{V_6}\int d^6 y \sqrt{h}\; R_{mn}(h)h^{mn}, \quad V_6 = \int d^6 y \sqrt{h}\,. 
\end{equation}
We see that $\sigma$ has a kinetic term and that the four-dimensional Newton's constant is
\begin{equation}
    \kappa_4^2 = \frac{\kappa^2_{10}}{V_6}\,.
\end{equation}
It will be convenient to define that
\begin{equation}\label{field-redif1}
    \Phi = \frac{\varphi}{2} - 6\sigma\,, \quad \Psi = \frac{\varphi}{2} + 2\sigma,
\end{equation}
from which it is straightforward to show that
\begin{equation}\label{field-redif2}
    -\frac{1}{2}\partial_\mu \Phi \partial^\mu \Phi -\frac{3}{2}\partial_\mu\Psi \partial^\mu \Psi = -\frac{1}{2}\partial_\mu \varphi \partial^\mu \varphi - 24 \partial_\mu \sigma \partial^\mu \sigma\,.
\end{equation}
Using this result, we can rewrite the gravity-scalar part of the four-dimensional action as
\begin{align}
    S = \frac{V_6}{2\kappa^2_{10}} \int d^4x \sqrt{-g}&\left[R(g) + e^{-4\sigma}\langle R(h)\rangle -\frac{1}{2}\partial_\mu \Phi \partial^\mu \Phi -\frac{3}{2}\partial_\mu\Psi \partial^\mu \Psi+\cdots\right]\,.
\end{align}

The ten-dimensional gauge field will give rise to a four-dimensional gauge field (but not necessarily with the same gauge group). The kinetic term for the 4-d gauge theory comes from 
\begin{align}
     -\frac{\kappa_{10}^2}{30g_{10}^2}\int d^{10}x \sqrt{-G} e^{-\varphi/2}\frac{1}{2}\text{tr} F_{ab}F^{ab} &\supset  -\frac{\kappa_{10}^2}{30g_{10}^2}\int d^{10}x \sqrt{-g}\sqrt{h} e^{-6\sigma-\varphi/2} e^{12\sigma} \frac{1}{2} \text{tr}F_{\mu\nu}F^{\mu\nu} \nonumber\\
     &= -\frac{\kappa_{10}^2 V_6}{30g_{10}^2}\int d^4 x \sqrt{-g} e^{6\sigma -\varphi/2}\frac{1}{2}\text{tr}F_{\mu\nu} F^{\mu\nu}\nonumber\\
     &= -\frac{\kappa_{10}^2 V_6}{30g_{10}^2}\int d^4 x \sqrt{-g} e^{-\Phi} \text{tr}|F_2|_4^2\,,
\end{align}
where we assumed the components $F_{\mu\nu}$ to depend solely on the external space $\left(x^\mu\right)$.

For the reduction of the $\Tilde{H}_3$ term, consider the decomposition
\begin{align}
    \frac{1}{2}\int e^{-\varphi}\Tilde{H}_3 \wedge * \Tilde{H_3} &= \frac{1}{2} \int d^{10}x \sqrt{-G} e^{-\varphi} \frac{1}{3!} \tilde{H}_{abc}\tilde{H}^{abc} \nonumber\\
    &= \frac{1}{12}\int d^{10} x \sqrt{-g}\sqrt{h} e^{-6\sigma -\varphi} \left[e^{18} \Tilde{H}_{\mu\nu\rho} \Tilde{H}^{\mu\nu \rho} + 3 e^{2\sigma} \tilde{H}_{\rho mn}\tilde{H}^{\rho mn} \right.\nonumber\\
    &\left. + \;3 e^{10\sigma} \tilde{H}_{\mu\nu p}\tilde{H}^{\mu\nu p} + e^{6\sigma} \tilde{H}_{mnp}\tilde{H}^{mnp}\right]\,.
\end{align}
The first term in the square bracket will give rise to the kinetic term for a two-form field in four dimensions, the second to scalar fields, the third to gauge fields, and the last vanishes if we assume $\Tilde{H}_3$ independent of the internal coordinates $y^m$. Assuming the $\tilde{H}_{\mu\nu\rho}$ components to be only $x^\mu$-dependent, we have
\begin{equation}
    \frac{1}{2}\int e^{-\varphi}\Tilde{H}_3 \wedge * \Tilde{H_3} \supset \frac{V_6}{12}\int d^4 x \sqrt{-g} e^{12 \sigma -\varphi} \Tilde{H}_{\mu\nu\rho}\Tilde{H}^{\mu\nu\rho}.
\end{equation}
This resembles the action for a two-form in four dimensions (although not canonically normalized), which can be dualized to a scalar field action. However, due to the modified Bianchi identity \eqref{Bianchiid}, this is not quite the action for a two-form in four dimensions. So, we dualize only after imposing \eqref{Bianchiid} as a constraint, i.e.,
\begin{equation}
    S\supset \frac{V_6}{2\kappa_{10}^2}\left\{-\frac{1}{2}\int e^{-2\Phi}\Tilde{H}_3\wedge *_4 \Tilde{H}_3 + \int a \left[d\Tilde{H}_3 - \frac{\kappa^2_{10}}{g^2_{10}}\left(\frac{1}{30}\text{tr}F_2\wedge F_2 - \text{tr}R_2 \wedge R_2\right)\right] \right\},
\end{equation}
where the integration is over the four-dimensional manifold. To get the dual scalar, we integrate out the three-form (see, e.g., \cite{Dasgupta:2008hb}). Varying with respect to $\Tilde{H}_3$ we find
\begin{equation}
    da = e^{-2\Phi}*_4 \Tilde{H}_3 \implies \Tilde{H}_3 = e^{2\Phi} *_4 da\;,
\end{equation}
and inserting this into the action again gives
\begin{equation}
    S\supset \frac{V_6}{2\kappa_{10}^2}\left\{-\frac{1}{2}\int e^{2\Phi} da \wedge *_4 da - \frac{\kappa^2_{10}}{g_{10}^2} \int a \left(\frac{1}{30}\text{tr}F_2\wedge F_2 - \text{tr}R_2 \wedge R_2\right)\right\},
\end{equation}
from which we can see that $a(x)$ has an axionic coupling with the gauge field, $a\epsilon^{\mu\nu\rho \sigma} F_{\mu\nu}F_{\rho\sigma}$.

Another axion comes from the term $H_{\rho m n} H^{\rho mn}$, because $H_{\rho mn} = \partial_\rho B_{mn}$:
\begin{equation}
    \frac{1}{2}\int e^{-\varphi}\Tilde{H}_3 \wedge * \Tilde{H_3} \supset \frac{V_6}{4}\int d^4 x \sqrt{-g} e^{-4 \sigma -\varphi} H_{\rho mn}H^{\rho mn} = \frac{V_6}{4}\int d^4 x \sqrt{-g} e^{-2\Psi} \partial_\rho B_{mn} \partial^{\rho}B^{mn}\;.
\end{equation}
The fact that $B_{mn}$ couples with $F_2\wedge F_2$ can be seen from the $B\wedge X_8$ coupling in the anomaly-induced action \eqref{anomalyterm} \cite{Witten:1984dg, Derendinger:1985cv, Ibanez:1986xy, Nilles:1997vk, Gukov:2003cy,Svrcek:2006yi}. This includes, for instance, the term
\begin{equation}
    \int B\wedge \text{tr} F_2^2\wedge \text{tr} F_2^2 \supset -\frac{1}{(2!)^5} \int d^{10}x \;\epsilon^{mn \mu\nu\rho\sigma pqrs}B_{mn} \text{tr}(F_{\mu\nu}F_{\rho \sigma}) \text{tr}(F_{pq}F_{rs}).
\end{equation}
Hence, if $F_{pq}(y)$ is non-vanishing in the internal manifold and is defined in the direction other than the two-cycle where $B_{mn}$ is defined, we can get a coupling of the form
\begin{equation}
    \int \epsilon^{mn}B_{mn} \text{tr}F_2 \wedge F_2,
\end{equation}
after the dimensional reduction. Defining $\epsilon^{mn}B_{mn} = 2\sqrt{3} b$, we have 
\begin{equation}
    S\supset  \frac{V_6}{2\kappa_{10}^2} \left(-\frac{3}{2}\int e^{-2\Psi} db\wedge *_4 db - \frac{\beta}{30} \int b \;\text{tr} F_2 \wedge F_2\right),
\end{equation}
where we collected all the numerical coefficients, including the values of the internal gauge field strengths and details of the internal manifold, into the dimensionfull quantity $\beta$. Using the definitions in \eqref{field-redif1} and using \eqref{field-redif2}, 
we can finally write the dimensionally reduced action as 
\begin{align}
    S = \frac{V_6}{2\kappa^2_{10}} \int d^4x \sqrt{-g}&\left[R(g) + e^{-4\sigma}\langle R(h)\rangle - \frac{\kappa^2_{10}}{30g^2_{10}}e^{-\Phi}\text{tr}|F_2|^2 -\frac{1}{2}\partial_\mu \Phi \partial^\mu \Phi - \frac{1}{2}e^{2\Phi} \partial_\mu a \partial^\mu a \right. \nonumber\\
    &\left. -\frac{3}{2}\partial_\mu\Psi \partial^\mu \Psi- \frac{3}{2} e^{-2\Psi} \partial_\mu b \partial^\mu b + \frac{1}{4}\left(\frac{\kappa_{10}^2}{g_{10}^2} a + \beta \;b \right)\epsilon^{\mu\nu\rho\sigma}\text{tr}\left( F_{\mu\nu}F^{\rho\sigma} \right) +  \cdots\right]\,.
\end{align}
The overall factor of $V_6/\kappa_{10}^2 = 1/\kappa_4^2$ is absorbed by the fields to make them dimensionfull, while the fiducial four-dimensional gauge coupling is $g_4 = g_{10}/\sqrt{V_6}$,
\begin{align}\label{4daction}
    S = \int d^4x \sqrt{-g}&\left[\frac{1}{2\kappa_4^2}R(g) + \frac{1}{2\kappa_4^2}e^{\kappa_4(\Phi- \Psi)/2} \langle R(h)\rangle - \frac{e^{-\kappa_4\Phi}}{4g^2_{4}}\frac{1}{30}\text{tr}(F_{\mu\nu}F^{\mu\nu}) -\right. \nonumber\\ 
    &\left. -\frac{1}{4}\partial_\mu \Phi \partial^\mu \Phi - \frac{1}{4}e^{2\kappa_4\Phi} \partial_\mu a \partial^\mu a 
    -\frac{3}{4}\partial_\mu\Psi \partial^\mu \Psi- \frac{3}{4} e^{-2\kappa_4\Psi} \partial_\mu b \partial^\mu b + \right. \nonumber\\
    & \left. + \frac{1}{8}\left(\frac{\kappa_4}{g_{4}^2} a + \frac{\beta}{\kappa_4} b\right)\epsilon^{\mu\nu\rho\sigma}\frac{1}{30}\text{tr}\left( F_{\mu\nu}F_{\rho\sigma} \right) + \cdots\right]\,.
\end{align}
Then, we find that $\Phi$ is the four-dimensional dilaton which fixes the physical four-dimensional gauge coupling
\begin{equation}
    g_{\text{YM}}^2 = \frac{g^{2}_{10}}{V_6} e^{\kappa_4\Phi_0 } = g_4^2e^{\kappa_4\Phi_0 }\,,
\end{equation}
while $\Psi$ is the four-dimensional moduli associated with the internal volume. Note the kinetic mixing between $\Phi$ and $a$, and between $\Psi$ and $b$. The parameter $\beta$ has the dimension of inverse mass squared. The dots represent many terms we have neglected, such as the runaway potential terms for $\Psi$ and $\Phi$ induced by fluxes \cite{Dine:1985rz, Becker:2002jj}. A more systematic way of looking at the four-dimensional dynamics is described in the next section.

%%%%%%%%%%%%%%%%%%%%%%%%%%%%%%%%%%%%%%%%  
\subsection{Four-dimensional action from supergravity}

The action \eqref{heteroticaction} is actually just the bosonic piece of the tree-level heterotic low-energy action, which also includes fermions. The theory is actually supersymmetric, with 16 supercharges. Supersymmetry helps control corrections to the theory and so it ensures that solutions to the supergravity equations are also solutions to the full string theory \cite{Candelas:1985en,Witten:1985xb}. Moreover, it helps to track the possible terms one can get after dimensional reduction. However, an arbitrary compactification would break supersymmetry completely, and these nice properties would be lost. In making contact with four-dimensional physics, one focuses on compactifications that preserve $\mathcal{N}=1$ supersymmetry in four-dimensions. This is the case if the internal space is a Ricci-flat, K\"ahler manifold with SU$(3)$ holonomy group, as it is for Calabi-Yau three-folds \cite{Candelas:1985en}.

Instead of diving into the details of compactification, we will start with a $\mathcal{N}=1$ supergravity model coupled with gauge and chiral superfields. In this case, the four-dimensional action is set by three ``functions'' of the superfields: the K\"ahler potential $K(T^I, \Bar{T}^{\Bar{J}})$, which gives the kinetic terms of the chiral fields, the gauge kinetic function $f_{ab}(T^I)$, which fixes the gauge fields kinetic terms, and $W(T^{I})$ which enters in the scalar potential. The gauge kinetic function is actually a set of functions, one for each component of the gauge group. Moreover, $K$ and $f_{ab}$ are holomorphic functions of the complex scalar part of the superfields, $\Bar{T}^I$. The bosonic part of the $\mathcal{N}=1$ action with vector and (neutral) chiral multiplets is \cite{Freedman:2012zz}
\begin{align}
    S_{\mathcal{N}=1} = \int d^4x \sqrt{-g} &\left[\frac{1}{2\kappa_4^2}R - K_{I\Bar{J}}\partial_\mu T^I \partial^\mu \Bar{T}^{\Bar{J}} - \frac{1}{4} \text{Re}(f_{ab}) F^{a}_{\mu\nu}F^{b\mu\nu} +\frac{1}{8}\text{Im}(f_{ab})\epsilon^{\mu\nu\sigma \rho}F_{\mu\nu}^a F_{\sigma \rho}^b - \right.\nonumber\\
    &\left.- e^{\kappa^2_4 K}\left(K^{I\Bar{J}}D_I W \overline{D_J W} -3\kappa^2_4 |W|^2\right)   \right]\;,
\end{align}
where 
\begin{equation}
    K_{I\Bar{J}} = \frac{\partial^2 K}{\partial T^I \partial \overline{T}^{\Bar{J}}}, \quad D_IW = \frac{\partial W}{\partial T^I} + \kappa^2_4 W \frac{\partial K}{\partial T^I}\;,
\end{equation}
and $K^{I\Bar{J}}$ is the inverse of $K_{I \Bar{J}}$.

Comparing the action just above with eq.~\eqref{4daction}, we find two chiral moduli $S = e^{-\kappa_4 \Phi} + i\kappa_4a$ and $T = e^{\kappa_4\Psi} +i\kappa_4 b$ and, from the scalar kinetic terms in \eqref{4daction}, we should have
\begin{equation}
    K = \kappa^{-2}_4 \ln (S+\Bar{S}) + \kappa^{-2}_4 3 \ln (T + \overline{T}),
\end{equation}
while, from the gauge kinetic term, 
\begin{equation}
     f_{ab} = \frac{1}{30}\delta_{ab}\left(\frac{S}{g_4^2}+\frac{\beta}{\kappa_4^2} T\right),
\end{equation}
where the indices are in the adjoint representation of the gauge group. The term dependent on $T$ in the gauge kinetic function was inferred from the axionic coupling of $b$ with $F\wedge F$. However, by supersymmetry, we know that Re$(T)$ should be coupled with $F\wedge * F$. So, the structure of supergravity action tells us that an extra term should be added to the action \eqref{4daction}:
\begin{align}
    S = \int d^4x \sqrt{-g}&\left[\frac{1}{2\kappa_4^2}R(g) -\frac{1}{4}\partial_\mu \Phi \partial^\mu \Phi-\frac{e^{2\kappa_4\Phi} }{4}\partial_\mu a \partial^\mu a 
        - \frac{1}{4}\left(\frac{e^{-\kappa_4\Phi}}{g^2_{4}}+ \frac{\beta}{\kappa_4^2}e^{\kappa_4 \Psi}\right)\frac{1}{30}\text{tr}(F_{\mu\nu}F^{\mu\nu})- \right.\nonumber\\
    &\left. -\frac{3}{4}\partial_\mu\Psi \partial^\mu \Psi- \frac{3}{4} e^{-2\kappa_4\Psi} \partial_\mu b \partial^\mu b + \frac{1}{8}\left(\frac{\kappa_4}{g_{4}^2} a + \frac{\beta}{\kappa_4} b\right)\epsilon^{\mu\nu\rho\sigma}\frac{1}{30}\text{tr}\left( F_{\mu\nu}F_{\rho\sigma} \right) + \cdots\right].
\end{align}

A constant superpotential is induced from $\Tilde{H}_3$ fluxes, but the expression for the potential is such that $W = W_0$ only generates runaway potentials\footnote{However, the so-called complex structure moduli can be stabilized by this effect (see e.g. \cite{Cicoli:2013rwa} and references therein). We are assuming this step was already done, such that only the dynamics of $S$ and $T$ matters.} for $S$ and $T$. Moreover, $S$ can be stabilized by gaugino condensation \cite{Affleck:1983rr, Affleck:1983mk, Affleck:1984xz, Dine:1985rz}. After $S$ stabilization, the gauge kinetic term will be canonically normalized, and the $a F\wedge F$ term will become a total derivative. So we can write
\begin{align}
    S = \int d^4x \sqrt{-g}&\left[\frac{1}{2\kappa_4^2}R(g) - \frac{1}{4}\left(\frac{1}{g^2_{\text{YM}}}+ \frac{\beta}{\kappa_4^2}e^{\kappa_4 \Psi}\right)\frac{1}{30}\text{tr}(F_{\mu\nu}F^{\mu\nu})- \right.\nonumber\\
    &\left. -\frac{3}{4}\partial_\mu\Psi \partial^\mu \Psi- \frac{3}{4} e^{-2\kappa_4\Psi} \partial_\mu b \partial^\mu b + \frac{1}{8} \frac{\beta}{\kappa_4} b\epsilon^{\mu\nu\rho\sigma}\frac{1}{30}\text{tr}\left( F_{\mu\nu}F_{\rho\sigma} \right) + \cdots\right].
\end{align}
Defining $\chi = \sqrt{3/2}\Psi$ and $\phi = \sqrt{3/2}b$, we finally have
\begin{align}\label{eq:final_4d_action}
    S = \int d^4x \sqrt{-g}&\left[\frac{1}{2\kappa_4^2}R(g) - \frac{1}{4g^2_{\text{YM}}}\frac{1}{30}\text{tr}(F_{\mu\nu}F^{\mu\nu})- \frac{1}{2}\partial_\mu\chi \partial^\mu \chi- \frac{1}{2} e^{-\sqrt{8/3}\kappa_4\chi} \partial_\mu \phi \partial^\mu \phi -\right. \nonumber\\
    &\left. -\frac{\beta}{4\kappa_4^2}e^{\sqrt{2/3}\kappa_4 \chi}\frac{1}{30}\text{tr}(F_{\mu\nu}F^{\mu\nu}) +\frac{\sqrt{2}}{8\sqrt{3}} \frac{\beta}{\kappa_4} \phi \epsilon^{\mu\nu\rho\sigma}\frac{1}{30}\text{tr}\left( F_{\mu\nu}F_{\rho\sigma} \right) + \cdots\right].
\end{align}

According to examples from \cite{Gukov:2003cy, Nilles:1997vk}, the parameter $\beta$ can be as large as $\mathcal{O}(10)$ in Planckian units.

%%%%%%%%%%%%%%%%%%%%%%%%%%%%%%%%%%%%%%%%  
\section{Dissipation terms in heterotic string compactification}
\label{section4}

Working with heterotic string compactification as described in the previous section, we consider an effective model of two scalar fields interacting with a non-Abelian gauge field. The Lagrangian density of the model is of the general form
\begin{eqnarray}
{\cal L} &=& \frac{M_{\rm Pl}^2}{2}R -\frac{1}{4}  F_{\mu \nu}^a F^{a,\mu \nu}-\frac{1}{2}\partial_\mu \chi\partial^\mu \chi
\nonumber \\
&-&\frac{e^{-\lambda_1 \chi/M_{\rm Pl}}}{2} \partial_\mu \phi \partial^\mu \phi
+ \lambda_2 e^{\lambda_3 \chi/M_{\rm Pl}} F_{\mu \nu}^a F^{a,\mu \nu} - \lambda_4 \frac{\phi}{M_{\rm Pl}} \epsilon^{\mu\nu\rho\sigma}
F_{\mu \nu}^a F_{\rho \sigma}^a
\nonumber \\
&-& V(\phi,\chi),
\label{lagr}
\end{eqnarray}
where $F_{\mu \nu}^a=\partial_\mu A_\nu^a - \partial_\nu A_\mu^a+ g f^{abc}A_\mu^b A_\nu^c$,
 $a\in \{1,\ldots,N^2-1\}$ for a $SU(N)$ gauge field, $\lambda_1,\ldots,\lambda_4$ are coupling constants
and $V(\phi,\chi)$ is the potential for the two scalar fields. 
Given a concrete compactification example, we can fix the $\lambda_i$ parameters. {}For example, comparing eq.~(\ref{lagr}) to \eqref{eq:final_4d_action}, we have for instance that
\begin{equation}
\lambda_1=2\sqrt{\frac{2}{3}},\quad
\lambda_2 = \frac{g_{\rm YM}^2 \bar \beta}{4},\quad 
\lambda_3=\sqrt{\frac{2}{3}},\quad \lambda_4 = \frac{g_{\rm YM}^2 \bar \beta}{4 \sqrt{6}},\quad 
g=g_{\rm YM}\sqrt{\frac{30}{N}},
\label{couplingsstring}
\end{equation}
with $\bar \beta = -\beta M_{\rm Pl}^2$ and $\kappa_4\equiv 1/M_{\rm Pl}$. In our numerical analysis in section~\ref{section5}, we consider both arbitrary couplings and the specific case in 
which the couplings are fixed to the heterotic model values given in eq.~(\ref{couplingsstring}). 

{}For the discussion below, we do not need to specify $V(\phi,\chi)$ or $N$ explicitly at the moment. 
However, for all numerical examples we take $N=3$ (i.e. an $SU(3)$ gauge sector), which could be thought of as part of the higher dimensional gauge group after symmetry breaking, such as the hidden $E_8$ sector.
Below, we will also assume that the scalar fields are homogeneous fields and, thus, are only time dependent: $\phi \equiv \phi(t)$ and $\chi\equiv \chi(t)$. Their coupling to the gauge field
$A_\mu$ will lead to the following contributions in the equation of motion for $\phi$ and $\chi$:
\begin{eqnarray}
&&\ddot \chi + 3 H \dot \chi + \frac{\lambda_1}{2 M_{\rm Pl}} e^{-\lambda_1 \chi/M_{\rm Pl}} \dot\phi^2 +V_{,\chi} 
-\frac{\lambda_2 \lambda_3 e^{\lambda_3 \chi/M_{\rm Pl}} }{M_{\rm Pl}} \langle  F_{\mu \nu}^a F^{a,\mu \nu}\rangle =0,
\label{eqchi}
\\
&&e^{-\lambda_1 \chi} \left( \ddot \phi + 3 H \dot \phi\right)
- \frac{\lambda_1 e^{-\lambda_1 \chi/M_{\rm Pl}}}{M_{\rm Pl}} \dot\phi \dot \chi + V_{,\phi} + \frac{\lambda_4}{M_{\rm Pl}}\epsilon^{\mu\nu\rho\sigma}
\langle F_{\mu \nu}^a F_{\rho \sigma}^a\rangle =0.
\label{eqphi}
\end{eqnarray}
As already shown in ref.~\cite{Laine:2021ego} and also discussed in ref. \cite{DeRocco:2021rzv}, dissipative processes involving the gauge particles tend to thermalize fast, with a rate  $\Gamma \sim 10 N^2 \alpha^2 T$, where $\alpha=g^2/(4\pi)$ is the Yang-Mills coupling, which is larger than the Hubble rate, i.e., $\Gamma>H$. The parameter space where this happens also turns out to be the one leading to the WI  regime\footnote{Note that the condition $\Gamma/H> 1$ implies that $\Gamma/H \sim 10 N^2 \alpha^2 T/H > 1$, or, equivalently, $T/H \gtrsim 1/(10 N^2 \alpha)$. In all of our numerical examples considered in the next section we used $N=3$ and $\alpha=0.1$. Thermalization in all of our examples then requires $T/H \gtrsim 1.1$. Note that WI itself is defined by having $T/H > 1$ and $T/H$ tends to increase in general in WI~\cite{Berera:2008ar,BasteroGil:2009ec, Kamali:2023lzq}. This ensures that $\Gamma/H > 1$ also holds throughout the WI numerical examples studied here. }. This allows us to treat the gauge field averages in (\ref{eqchi}) and (\ref{eqphi}) as ensemble averages over an approximated equilibrium state, which also holds true if the scalar fields are slowly
moving. The field averages can then be viewed as describing the response of the system to the small time
variation of the scalar fields and we can use a standard linear response theory expansion for them~\cite{Berera:2004kc},
such that
\begin{align}\label{ave1}
\langle \epsilon^{\mu\nu\rho\sigma} &F_{\mu \nu}^a F_{\rho \sigma}^a\rangle \simeq \langle \epsilon^{\mu\nu\rho\sigma}F_{\mu \nu}^a F_{\rho \sigma}^a\rangle_0 \,+ \\
& i \frac{\lambda_4}{M_{\rm Pl}} \int_0^t dt' \int d^3 x' \phi(t') \langle [\epsilon^{\mu\nu\rho\sigma}F_{\mu \nu}^a({\bf x},t)  F_{\rho \sigma}^a({\bf x},t) ,
\epsilon^{\mu'\nu'\rho'\sigma'}F_{\mu' \nu'}^b({\bf x}',t')  F_{\rho' \sigma'}^b({\bf x}',t')]\rangle_0\nonumber
\end{align}
and
\begin{align}\label{ave2}
\langle F_{\mu \nu}^a F_{\mu \nu}^a\rangle \simeq \langle  & F_{\mu \nu}^a F_{\mu \nu}^a\rangle_0 \,- \\
& i \lambda_2 \int_0^t dt' \int d^3 x' e^{\lambda_3 \chi(t')} \langle [F_{\mu \nu}^a({\bf x},t)  F^{a,\mu \nu}({\bf x},t) \,,
F_{\mu' \nu'}^b({\bf x}',t')  F^{b,\mu' \nu'}({\bf x}',t')]\rangle_0\,, \nonumber
\end{align}
where $\langle \ldots \rangle_0$ denotes averages over the thermal equilibrium state. Note that the local thermal equilibrium terms will generally contribute to thermal corrections to the effective potential for the $\phi$ and $\chi$ background fields. The local thermal equilibrium term in (\ref{ave1}), since it is a Chern-Simons term, gives no local thermal contribution in (\ref{eqphi}), since it vanishes identically (note, however, that, as in the axion case, nonperturbative contributions can still generate a thermal mass term, but this is
highly suppressed~\cite{Berghaus:2019whh}). However, the local thermal contribution in (\ref{ave2}) does not vanish and must be considered. We can associate it with the calculation of the thermodynamic potential performed in the pure gauge field case~\cite{Andersen:2002ey}, with the leading order contribution in the gauge coupling, ${\cal O}(g^2)$, given by\footnote{We note that $\langle  F_{\mu \nu}^a F^{a,\mu \nu}\rangle$ is related to the trace of the energy-momentum tensor  and to the beta function for the gauge coupling, $\beta(g)$, through $\langle T^\mu_\mu \rangle = T^5 \frac{\partial}{\partial T} \left( \frac{P}{T^4} \right) = \frac{\beta(g)}{2g} \langle F^a_{\mu\nu} F^{a\mu\nu} \rangle$, where $P=-\Omega$ and $\Omega$ is the thermodynamic potential (see, e.g. \cite{Andersen:2002ey}).}
\begin{equation}
\Delta V_{\rm eff}(\chi,T) = -\lambda_2 e^{\lambda_3 \chi/M_{\rm Pl}}  \langle  F_{\mu \nu}^a F^{a,\mu \nu}\rangle_0 \simeq 
\lambda_2 e^{\lambda_3 \chi/M_{\rm Pl}} N(N^2-1) \frac{g^2 T^4}{36}\,.
\label{DeltaVeff}
\end{equation}
In this case, the total energy density will be given by
\begin{equation}
\rho_T = e^{-\lambda_1 \chi/M_{\rm Pl}} \frac{\dot{\phi}^2}{2} + \frac{\dot\chi^2}{2} +  V_{\rm eff}(\phi,\chi,T)  + T s\,,
\label{rho}
\end{equation}
where $s$ is the entropy density, defined in terms of the effective potential as
\begin{equation}
s= - \frac{\partial V_{\rm eff}(\phi,\chi,T)}{\partial T},
\label{ent}
\end{equation}
with
\begin{equation}
V_{\rm eff}(\phi,\chi,T) = V(\phi)+V(\chi) - \frac{\pi^2 g_*}{90}T^4+\lambda_2 e^{\lambda_3 \chi/M_{\rm Pl}} N(N^2-1) \frac{g^2 T^4}{36}\,.
\label{Veffchi}
\end{equation}
and from eq.~(\ref{ent}),
\begin{eqnarray}
s= \frac{2 \pi^2 g_*}{45}T^3
- \lambda_2 e^{\lambda_3 \chi/M_{\rm Pl}} N(N^2-1) \frac{g^2 T^3}{9}\,,
\end{eqnarray}
In the above equations, we have also explicitly included the ideal gas contribution from the gauge fields, with $g_*= 2 (N^2-1)$, i.e., we are assuming that only the gauge field contributes for the thermal bath degrees of freedom. 
Then, the total energy density (\ref{rho}) is given by
\begin{eqnarray}
\!\!\!\!\!\!\!\!\!\rho_T &=& e^{-\lambda_1 \chi/M_{\rm Pl}} \frac{\dot{\phi}^2}{2} + \frac{\dot\chi^2}{2} + V(\phi) +V(\chi) + \frac{\pi^2 g_*}{30}T^4
- \lambda_2 e^{\lambda_3 \chi/M_{\rm Pl}} N(N^2-1) \frac{g^2 T^4}{12}.
\label{rhoT}
\end{eqnarray}

The nonlocal terms given by the last terms in (\ref{ave1}) and (\ref{ave2}), on the other hand, are both nonvanishing and will 
lead explicitly to dissipation terms, $\propto \dot \phi$ and $\dot \chi$, when expanding the fields close to equilibrium~\cite{Gleiser:1993ea, Berera:1998gx, Berera:2001gs, Berera:2004kc, Berera:2008ar}.
However, solving explicitly for the thermal averages for the gauge fields in eqs.~(\ref{ave1}) and (\ref{ave2}) is quite
cumbersome and the results are only known numerically~\cite{Moore:2010jd}. 
Based on the results in \cite{Moore:2010jd, Laine:2016hma}, we have, for instance, that the second term on the right-hand side of eq.~(\ref{ave1}) that contributes to dissipation in the equation of motion for $\phi$ gives~\cite{Laine:2021ego})
\begin{equation}
\frac{\lambda_4}{M_{\rm Pl}}\langle \epsilon^{\mu\nu\rho\sigma} F_{\mu \nu}^a F_{\rho \sigma}^a\rangle_{\rm diss} \sim \Upsilon_\phi(T) \dot{\phi}\,,
\label{dissphi}
\end{equation}
where
\begin{equation}
\Upsilon_\phi(T)= \kappa \frac{T^3}{M_{\rm Pl}^2}\,.
\label{Upsilonphi}
\end{equation}
The coefficient $\kappa$ is given by
\begin{equation}
\kappa \simeq 1.2 \frac{\lambda_4^2 (g^2 N)^3 (N^2-1)}{\pi}\left[\ln\left(\frac{m_D}{\gamma}\right)+3.041\right]\,,
\label{kappa}
\end{equation}
where $m_D^2=g^2 N T^2/3$ is the Debye mass squared of the Yang-Mills plasma and $\gamma$ is given by the solution of
\begin{equation}
\gamma=\frac{g^2 N T}{4 \pi} \left[\ln\left(\frac{m_D}{\gamma}\right)+3.041\right]\,.
\end{equation}
Likewise, in the equation of motion for $\chi$,  the second term in right-hand side in eq.~(\ref{ave2}) that contributes to dissipation gives\footnote{Note that in this case, the computation for the second term in (\ref{ave2}) which leads to dissipative effects, can be related to the calculation of the bulk viscosity coefficient in gauge theory. The result of this calculation is, unfortunately, only known for the case of QCD~\cite{Arnold:2006fz}, i.e., for the case of $SU(3)$.}~\cite{Laine:2010cq}
\begin{equation}
-\frac{\lambda_2 \lambda_3 e^{\lambda_3 \chi/M_{\rm Pl}} }{M_{\rm Pl}} \langle  F_{\mu \nu}^a F^{a,\mu \nu}\rangle_{\rm diss}
\sim  \Upsilon_\chi(T) \, \dot\chi\, e^{2 \lambda_3 \chi/M_{\rm Pl}}\,,
\label{disschi}
\end{equation}
where
\begin{equation}
\Upsilon_\chi(T) \sim (\lambda_2\lambda_3)^2 \frac{(12 \pi \alpha)^2}{\ln(1/\alpha)} \frac{T^3}{M_{\rm Pl}^2}\,.
\label{Upsilonchi}
\end{equation}

Note that the gauge-field production allowed by the axial coupling $\phi F \tilde{F}$ (see e.g. ref.~\cite{Barnaby:2011vw}) cannot invalidate the thermal description used to compute $\Upsilon_{\phi}(T)$. This is because of the thermal state considered, over which the gauge fields acquire a Debye mass $\sim gT$ and undergo rapid Landau damping in the plasma; these effects erase coherent tachyonic growth long before it can compete with thermal scatterings. 

Including the dissipation terms given by (\ref{dissphi}) and (\ref{disschi}) in (\ref{eqchi}) and
(\ref{eqphi}) and also taking into account the thermal contribution (\ref{DeltaVeff}), we finally find that the effective equations of motion for $\chi$ and $\phi$ are given by
\begin{eqnarray}
&&\ddot \chi + 3 H \dot \chi + \frac{\lambda_1}{2 M_{\rm Pl}} e^{-\lambda_1 \chi/M_{\rm Pl}} \dot\phi^2 +V_{,\chi} + \frac{\lambda_2 \lambda_3}{M_{\rm Pl}} e^{\lambda_3 \chi/M_{\rm Pl}} N(N^2-1) \frac{g^2 T^4}{36 } +\Upsilon_\chi(T) \, \dot\chi\, e^{2 \lambda_3 \chi/M_{\rm Pl}} =0\,,
\nonumber\\
\label{eomchi}
\\
&&e^{-\lambda_1 \chi/M_{\rm Pl}} \left( \ddot \phi + 3 H \dot \phi\right)
- \frac{\lambda_1 e^{-\lambda_1 \chi/M_{\rm Pl}}}{M_{\rm Pl}} \dot\phi \dot \chi + V_{,\phi} +\Upsilon_\phi(T) \,\dot\phi =0\,.
\label{eomphi}
\end{eqnarray}
The evolution of the system is then fully determined once the thermal bath and the first {}Friedmann equation are also considered, which are given explicitly as follows:
\begin{subequations}
\begin{align}
&T \dot s + 3H T s  - \Upsilon_\chi (T) e^{2\lambda_3 \chi/M_{\rm Pl}}\dot{\chi}^2 - \Upsilon_\phi \dot{\phi}^2 =0\,,
\label{entropyeq}\\
& 3H^2M^2_{\rm Pl} = e^{-\lambda_1 \chi/M_{\rm Pl}} \frac{\dot{\phi}^2}{2} + \frac{\dot\chi^2}{2} + V(\phi) +V(\chi) + \frac{\pi^2 g_*}{30}T^4
- \lambda_2 e^{\lambda_3 \chi/M_{\rm Pl}} N(N^2-1) \frac{g^2 T^4}{12}\,.
\end{align}
\end{subequations}
Note that the entropy equation~\eqref{entropyeq} can also be interpreted as a differential equation that governs the coupled evolution of the temperature 
and the fields, with the dissipation terms acting as sources of entropy production.

%%%%%%%%%%%%%%%%%%%%%%%%%%%%%%%%%%
\section{Numerical results for the background equations}
\label{section5}

\begin{figure}[htb!]
     \centering
         \includegraphics[width=1.0\textwidth]{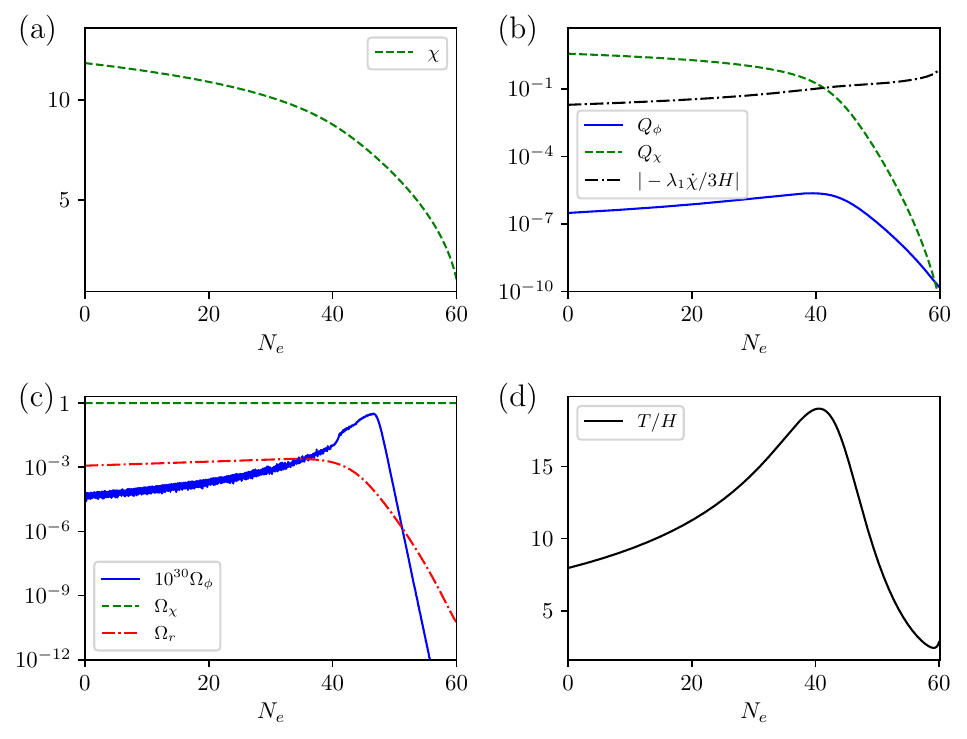}
            \caption{Near single-field scenario with kinetic coupling. The chosen parameters were $m_{\phi} = m_{\chi} = 10^{-6} \Mp$, $N=3$, $\alpha = 0.1$, and $\beta = 30$ (see definitions in eqs.~\eqref{couplingsstring}). The initial conditions are given by $(\chi_0, \chi'_0, \phi_0, \phi'_0, T_0, s_0)=(11.85, -0.04, 6.85\times 10^{-15}, -2.41\times 10^{-12}, 3.86 \times 10^{-5}, 2.89\times 10^{-9})$, in $ \Mp$ units. Both $\phi$ and $\Omega_\phi$ remain vanishingly small throughout the evolution. Here, $-\lambda_1 \dot{\chi} > 0$, signaling a flow of energy from $\phi$ to $\chi$ due to the kinetic coupling.}
        \label{fig:het}
\end{figure}

\begin{figure}[htb!]
     \centering
         \includegraphics[width=1.0\textwidth]{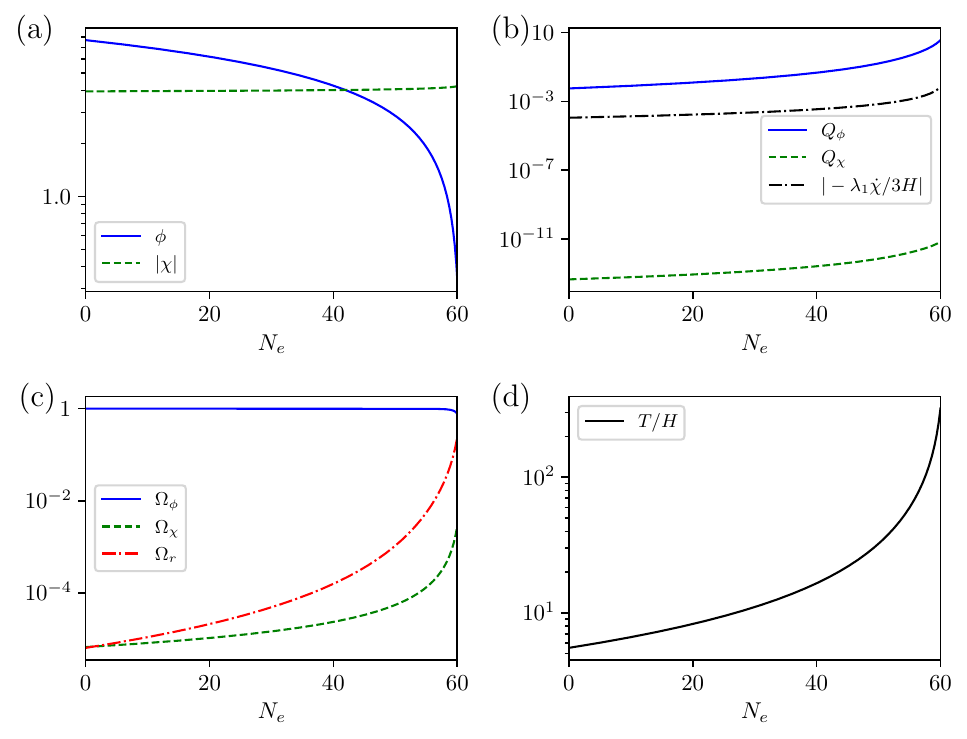}
            \caption{Minimal WI–like scenario. The chosen parameters were $m_{\phi} =  2\times 10^{-5} \Mp$, $m_{\chi} = 10^{-7} \Mp$, $N=3$, $\alpha = 0.1$, and $\lambda_i = (0.34, -2.47, 2.67, 7.0)$. The initial conditions are $(\chi_0, \chi_0', \phi_0, \phi'_0, T_0, s_0)=(-3.93, 0.0, 7.68, -0.07, 3.47 \times 10^{-4}, 2.93\times 10^{-10})$, in $ \Mp$ units. Here, $-\lambda_1 \dot{\chi} > 0$, signaling a flow of energy from $\phi$ to $\chi$ due to the kinetic coupling.}
        \label{fig:phidom}
\end{figure}

\begin{figure}[htb!]
     \centering
         \includegraphics[width=1.0\textwidth]{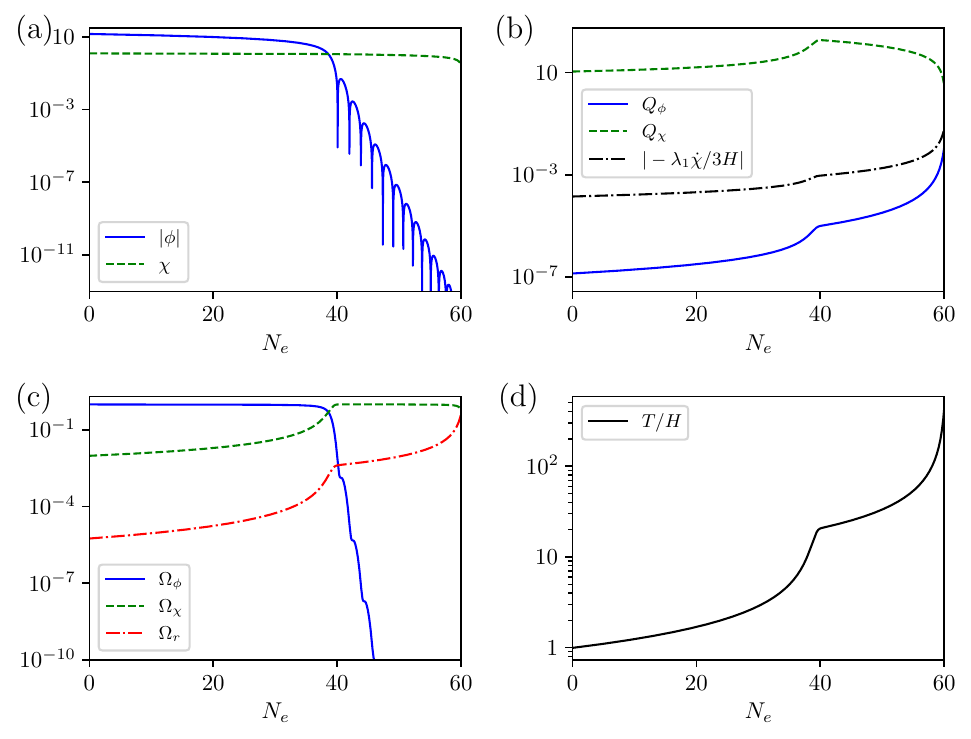}
            \caption{String-inspired model with crossover and strong $\chi$--driven dissipation. The chosen parameters were $m_{\phi} = 3.36 \times 10^{-6} \Mp$, $m_{\chi} = 3.91\times 10^{-6} \Mp$, $N=3$, $\alpha = 0.1$, and $\lambda_i = (0.29, -4.54, 6.73, 1.42)$. The initial conditions are $(\chi_0, \chi_0', \phi_0, \phi'_0, T_0, s_0) = (1.23, -0.001, -14.40, 0.20, 1.97\times 10^{-5}, 4.48\times 10^{-10})$, in $ \Mp$ units. Here, $-\lambda_1 \dot{\chi} > 0$, signaling a flow of energy from $\phi$ to $\chi$ due to the kinetic coupling.}
        \label{fig:cross1}
\end{figure}

\begin{figure}[htb!]
     \centering
         \includegraphics[width=1.0\textwidth]{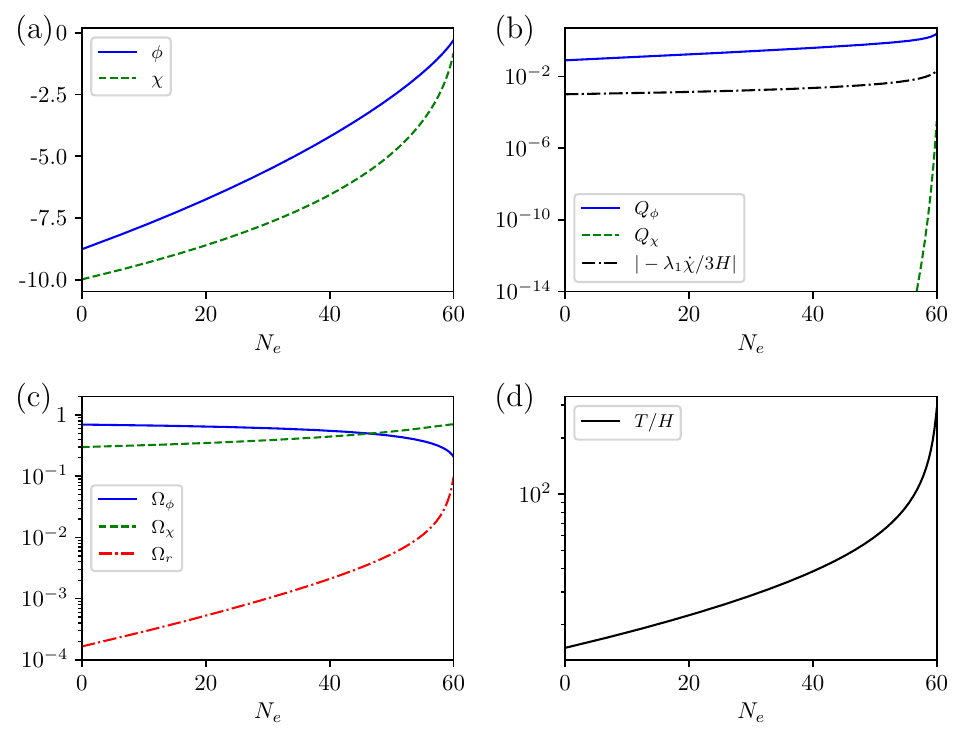}
            \caption{Crossover scenario with $\phi$--driven dissipation. The chosen parameters were $m_{\phi} = 10^{-5} \Mp$, $m_{\chi} = 5.75 \times 10^{-6} \Mp$, $N=3$, $\alpha = 0.1$, and $\lambda_i = \{ 0.05, -3.29, 5.0, 8.56 \}$. The initial conditions are $(\chi_0, \chi_0', \phi_0, \phi'_0, T_0, s_0) = (-9.80, 0.06, -8.86, 0.09, 6.62\times 10^{-4}, 2.04\times 10^{-9})$, in $ \Mp$ units. Here, $-\lambda_1 \dot{\chi} < 0$, so the kinetic coupling acts as a driving term for $\phi$.}
        \label{fig:cross2}
\end{figure}

After deriving the set of differential equations that govern the background evolution of our system, we probe the solutions that arise from different regions of parameter space and initial conditions. As expected, an analytical approach is somewhat elusive even under the slow-roll approximation, the natural exception being the standard case where the system effectively behaves as single-field with a radiation bath. Thus, we have turned to numerical methods, which reveal the rich variety of behaviors shown in figures~\ref{fig:het}-\ref{fig:cross2}. 
{}For the sake of concreteness, we have considered two quadratic potentials, i.e., 
\begin{equation}
    V(\phi) = \frac{1}{2}m_{\phi}^2 \phi^2\,, \qquad V(\chi) = \frac{1}{2}m_{\chi}^2 \chi^2\,.
\end{equation}
One can think of these as expansions of the fields around some local minima. The reason behind choosing such Gaussian potentials is that we want to emphasize the generic features of this model---the kinetic coupling that gets generated naturally in string theory---rather than focus on specific characteristics of fine-tuned potentials. 

Mechanisms for moduli stabilization that fixes the masses have been well studied in the literature (see \cite{Cicoli:2023opf} and references therein). Since we are interested in the WI phenomenology of the Lagrangian \eqref{lagr}, we solved the corresponding equations of motion numerically for a range of masses and $\{\lambda_i\}$, which allow us to find parameter regions that lead to successful WI regimes. Although string-inspired, all the numerical values of the parameters $\{\lambda_i\}$ considered are within the same order of magnitude obtained from concrete string theory compactification examples \cite{Gukov:2003cy}.

To assess the viability of different inflationary trajectories, we performed a numerical scan of $2 \times 10^4$ simulations using the LSODA solver~\cite{LSODA} with relative and absolute tolerances of $10^{-8}$, varying model parameters and initial conditions across a broad region of the parameter space. The couplings $\lambda_i$ ($i=1,\ldots,4$) were sampled log-uniformly in magnitude over $[10^{-6}, 10^4]$ with random sign, the scalar masses $m_\phi, m_\chi$ log-uniformly over $[10^{-7}, 10^{-3}]\,M_{\rm Pl}$, initial field values $\phi_0, \chi_0$ uniformly in $[-20, 20]\,M_{\rm Pl}$, initial velocities $d\phi/dN, d\chi/dN \in [-0.2, 0.2]\,M_{\rm Pl}$, and initial radiation energy density $\rho_0$ log-uniformly in $[10^{-15}, 10^{-8}]\,M_{\rm Pl}^4$, with $N=3$ and $\alpha=0.1$ throughout. A run is classified as viable inflation if it achieves at least 60 e-folds before $\epsilon_H = -\dot H/H^2\to 1$; WI additionally requires $T/H > 1$ for more than $90\%$ of the last 60 e-folds. 
We define physically valid configurations as those for which the background evolution remains numerically stable and all energy densities remain positive throughout the integration.
Of the physically valid configurations, approximately $800$ satisfied the inflation criterion, of which $\sim 380$ sustained WI throughout. Among these, roughly half 
correspond to effectively minimal WI, in the sense that the inter-field couplings remain perturbatively small ($|\lambda_{1,2,3}|<1$), so that the dynamics is dominated by a single field.
The other half exhibit genuine two-field dynamics with at least one coupling at $\mathcal{O}(1)$ or above, including configurations where $\chi$ contributes significantly to the energy budget. In $\sim 81\%$ of all viable warm runs the axion dominates at the end of inflation, confirming that $\phi$ generically drives the accelerated expansion. 
This provides quantitative support for the conclusion that warm inflation is generically driven by the axion field in this class of models, while dilaton-driven WI is comparatively disfavored.
The strong dissipative regime ($Q_\phi > 1$) is reached in roughly $88\%$ of the warm solutions. 
No statistically significant preference 
for the sign of any $\lambda_i$ was observed (including for the negative $\lambda_2$ values natural to the heterotic construction), indicating that the 
realization of WI is robust across this string-inspired parameter space.

%%%%%%%%%%%%%%%%%%%%%%%%%%%%%%%%%%%%%%%%%%%

%%%%%%%%%%%%%%%%%%%%%%%%%%%%%%%%%%%%%%%%%%%%%%%%%%

In figures~\ref{fig:het}--\ref{fig:cross2} we present four representative realizations of the model. In each case, panel (a) displays the evolution of the background fields during the last 60 e-folds of inflation. Panel (b) quantifies the relative importance of the dissipative coefficients and of the kinetic coupling in comparison with the Hubble friction term, through the ratios $Q_i \equiv \Upsilon_i/(3H)$. Panel (c) shows the evolution of the corresponding energy densities, while panel (d) illustrates the ratio $T/H$, which serves as an indicator of the onset and persistence of the WI regime. The specific initial conditions and parameter choices are indicated in each figure. We have chosen solutions with $T/H > 1$.
{}For the chosen parameters ($N=3$, $\alpha=0.1$), this condition also implies $\Gamma/H \gtrsim 1$, ensuring that the system remains close to thermal equilibrium and justifying the use of the linear-response approximation.
The time variable is the number of e-folds in every case. 
The parameter scan was performed using the LSODA solver, while the representative benchmark solutions shown in the figures were recomputed using \texttt{solve\_ivp} routine from \texttt{SciPy} 
for consistency checks and visualization.
The \texttt{solve\_ivp} routine uses as default an explicit Runge--Kutta method of order 5(4) with adaptive step size control. The relative and absolute tolerances were set to $10^{-12}$. Finally, the integration was terminated using an event function implementing the end-of-inflation condition $\epsilon_H= 1$.

We begin with figure~\ref{fig:het}, which depicts a realization of the (true) heterotic model defined by eqs.~\eqref{eq:final_4d_action} and~\eqref{couplingsstring}. In practice, this configuration behaves effectively as a single-field model. Its main distinction from minimal WI arises from the presence of the kinetic coupling, which becomes dominant during approximately the last third of the evolution, as shown in panel (b). Throughout most of inflation, the field $\phi$ remains effectively frozen near the origin, and its associated dissipative coefficient becomes comparable to that of $\chi$ only close to the end of inflation. The late-time dominance of the kinetic coupling appears to hinder a smooth transition to a radiation-dominated era.
Here we also find that unless the coupling parameter $\bar \beta$ in eq.~(\ref{couplingsstring}) is sufficient
large, $\bar \beta \gtrsim 30$, sufficient inflation cannot happen throughout in the warm regime (recalling that for the chosen parameters, thermalization restricts inflation to be in the warm regime throughout the dynamics).

{}Figure~\ref{fig:phidom} illustrates a more conventional WI scenario, in which a single field, $\phi$, drives the accelerated expansion. In this case, the dissipative dynamics is likewise dominated by $\phi$, as indicated by the behavior of $Q_\phi$ in panel (b), and the kinetic coupling remains subdominant throughout the evolution. As a result, the system undergoes a smooth transition to radiation domination. It is worth emphasizing, however, that this setup should be regarded as string-inspired rather than derived directly from the heterotic construction.

In another string-inspired realization, displayed in figure~\ref{fig:cross1}, the two fields successively dominate distinct stages of the inflationary evolution, as can be seen from the energy densities shown in panel (c). 
Despite this exchange in the dominant contribution to the energy budget, the dissipative dynamics remains governed by $\chi$ throughout the entire evolution, including during the interval in which $\phi$ dominates the total energy density. 
The system evolves in a strong dissipative regime at all times.
The transition between the $\phi$-dominated and $\chi$-dominated phases generates a pronounced feature around $N_e \approx 40$, which is reflected in several background quantities. 
In particular, the slow-roll parameter $\epsilon_H$ exhibits a sharp decrease at this point, signaling a transient departure from standard slow-roll behavior. 
Such non-trivial background dynamics may imprint distinctive signatures on the primordial perturbations, an issue that we defer to future work where a dedicated analysis of the fluctuation sector will be carried out.

{}Finally, figure~\ref{fig:cross2} presents a scenario in which the two fields again dominate at different times, albeit through a smoother transition than in the previous case. In contrast to figure~\ref{fig:cross1}, the dissipative dynamics remains predominantly controlled by $\phi$ throughout the evolution.

Note that the sign of the kinetic mixing term is controlled by $-\lambda_1 \dot{\chi}$. When $-\lambda_1 \dot{\chi} > 0$, the coupling acts as an effective friction term for $\phi$ (energy flows from $\phi$ to $\chi$), while for $-\lambda_1 \dot{\chi} < 0$ it acts as a driving term, sourcing energy into $\phi$. 

%%%%%%%%%%%%%%%%%%%%%%%%%%%%%%%%%%%%%%%%
\section{Comments on the cosmological perturbations
for the heterotic-like two-scalar field model}
\label{sec6}

Although our analysis has focused on the background dynamics, it is useful to briefly 
comment on the expected behavior of cosmological perturbations in the WI setup. 
Perturbations in two-field WI models have been investigated in the 
literature~\cite{Li:2019zbk, Wang:2018cev}, although typically in scenarios where the 
scalar fields interact only through the potential or evolve independently, each 
potentially driving a phase of WI.
As in multi-field cold inflation models, one expects two types of scalar perturbations: 
the curvature perturbation $\delta\sigma$ and an isocurvature mode $\delta s$, 
which are related to the field fluctuations $(\delta\phi,\delta\chi)$ through a 
rotation in field space. Existing analyses (e.g.~\cite{Li:2019zbk}) indicate that, 
in many cases, the curvature perturbation dominates the final spectrum. A full treatment 
of perturbations in the present model, which includes non-trivial kinetic structure and 
dissipation, would require a dedicated analysis and is beyond the scope of this work.
Nevertheless, our results show that, for appropriate parameter choices---such as in the 
heterotic-inspired benchmark of figure~\ref{fig:het}---the dynamics is effectively single-field, 
being dominated by the $\chi$ field while $\phi$ remains subleading. In this regime, 
it is reasonable to approximate the perturbations by those of a single-field WI model. 
The curvature power spectrum can then be written in the form~\cite{Ramos:2013nsa,Kamali:2023lzq}
\begin{equation}\label{power_spectra}
P_{R}\simeq \left(\frac{H^2}{2\pi\dot{\chi}}\right)^2
\left(1+2n_{\star}+\frac{2\sqrt{3}Q}{\sqrt{3+4\pi Q}}\frac{T}{H}\right)
G(Q)\big|_{k_{\star}=a_\star H_\star},
\end{equation}
where $n_\star$ corresponds to the inflaton occupation number. In the case of thermalized inflaton fluctuations, 
$n_\star=n_{BE}\equiv 1/(e^{H/T}-1)$, while $n_\star=0$ in the absence of thermalization. 
The realization of either case depends on the efficiency of scattering processes involving the inflaton 
and the thermal bath. The function $G(Q)$ encodes the effects of dissipation and the coupled dynamics of 
inflaton and radiation perturbations (see, e.g.,~\cite{Rodrigues:2025neh} for a recent discussion).
In this effective single-field regime, the standard WI expressions can be used to estimate 
observable quantities such as the tensor-to-scalar ratio $r$ and the spectral tilt 
$n_s$~\cite{Ramos:2013nsa,Kamali:2023lzq,Rodrigues:2025neh}. {}For the benchmark shown in figure~\ref{fig:het}, 
characterized by dynamics dominated by the $\chi$ field with a quadratic potential $V(\chi)=m_\chi^2\chi^2/2$ 
and a dissipation coefficient $\Upsilon\propto T^3$, we obtain\footnote{These results can be readily obtained 
using the public code \texttt{WI2easy}~\cite{Rodrigues:2025neh}.}
$(r,n_s)\sim (0.01,\,0.98)$,
at a dissipation ratio $Q \sim 10^{-4}$. Although this corresponds to the weak dissipative regime ($Q \ll 1$), 
the system still satisfies the warm inflation condition $T/H > 1$; for this benchmark we find $T/H \sim 7$, 
ensuring consistency with the WI framework.
As in standard WI, the tensor-to-scalar ratio is suppressed relative to its cold inflation counterpart. 
Parametrically,
\begin{equation}
r \sim \frac{16\epsilon}{(1+Q)^2}\frac{H}{T},
\end{equation}
so that both dissipation ($Q>0$) and the presence of a thermal bath ($T/H>1$) contribute to reducing $r$. 
In the present case, this suppression is moderate due to the small value of $Q$, but it can become significantly 
stronger in the regime $Q \gtrsim 1$.
While the obtained value of $r$ is consistent with current observational bounds ($r<0.036$~\cite{BICEP:2021xfz}), 
the spectral tilt lies slightly outside the $2\sigma$ region from Planck. 
We emphasize, however, that these estimates may change when both scalar fields contribute non-trivially to the 
dynamics, or for different choices of potentials motivated by string theory (see, e.g.,~\cite{Chakraborty:2025yms,Chakraborty:2026eep} 
for one-field examples).

%%%%%%%%%%%%%%%%%%%%%%%%%%%%%%%%%%%%%%%%  
\section{Conclusion}
\label{conclusions}

In this paper, we have derived a model of WI from heterotic string theory. WI provides a compelling alternative to standard cold inflation by incorporating dissipative effects that sustain a thermal bath during the epoch of accelerated expansion. In the context of heterotic string theory, we show that WI can be naturally realized due to the presence of moduli fields (such as the axio-dilaton) as well as a generic kinetic coupling between the various fields. The presence of a four-dimensional gauge field, after compactification, provides the necessary ingredient for a radiation bath, which is also kinetically coupled to the dilaton. The strength of this approach lies in not requiring one to fine-tune specific couplings in the potential terms of the various fields, and hence our findings are generic to WI models from heterotic string theory and do not depend sensitively on the potential.

The model consists of a non-Abelian gauge sector coupled to two scalar fields: 
an axion-like pseudoscalar $\phi$ and a dilaton $\chi$, with a non-trivial kinetic mixing between them. A key outcome of 
our analysis is the distinct dynamical role played by these two fields. The axion field $\phi$, protected by a shift symmetry in 
its coupling to the gauge sector, is largely insensitive to thermal corrections and can naturally sustain WI. 
In contrast, the dilaton $\chi$ couples exponentially to the gauge kinetic term, $e^{-\lambda_3 \chi}\,{\rm Tr}(F_{\mu\nu}F^{\mu\nu})$, 
and thus generically acquires significant thermal corrections that tend to disrupt the slow-roll evolution  in WI\footnote{The first study of the disruptive effects of thermal corrections in WI was done in ref.~\cite{Yokoyama:1998ju}. A dynamical system analysis in WI demonstrating that thermal corrections to the inflaton potential disrupt the inflationary attractor trajectory is presented in ref.~\cite{Moss:2008yb}. A somewhat similar situation to the one studied in this paper has also been shown to occur in warm chromoinflation~\cite{Kamali:2024qme}, where the presence of a thermal mass for the gauge field background was shown to make the gauge field condensate unstable and to vanish.}.

Our numerical analysis of the coupled system across a broad parameter space reveals a variety of dynamical regimes, 
including effectively single-field, multi-field, and crossover solutions. We find that WI is typically realized 
along the axion direction, while sustained dilaton-driven WI is disfavored over most of the parameter space due 
to thermal backreaction effects. Although solutions exist in which the dilaton temporarily dominates the dynamics 
or becomes relevant near the end of inflation, the majority of viable configurations exhibit effectively 
single-field behavior driven by the axion. These results indicate that, within heterotic-inspired constructions, 
thermal effects provide a dynamical mechanism that favors axion-driven WI while limiting the role of the dilaton.

We also identify regions of parameter space in which one of the scalar fields can be effectively neglected, 
leading to dynamics closely resembling minimal WI\footnote{See \cite{Das:2019acf} for discussions on the consistency of minimal WI and the Swampland program.}. In addition, we find that even when starting from initial 
conditions with negligible dissipation, the structure of the couplings generically drives the system toward a 
warm inflationary regime. The interaction parameters $\lambda_i$ $(i=1,\ldots,4)$ are taken within ranges 
consistent with heterotic constructions, typically $\mathcal{O}(1\text{--}10)$, ensuring that the model 
remains well motivated from a fundamental perspective.

A comment on field excursions is in order. The physically relevant quantity for the swampland
distance conjecture~\cite{Ooguri:2006in} is the geodesic distance in field space, which in our
model is measured by the metric $\mathrm{d}s^2 = \mathrm{d}\chi^2 +
e^{-\lambda_1\chi/M_{\rm Pl}}\,\mathrm{d}\varphi^2$. Due to the non-trivial kinetic coupling,
this proper distance can differ significantly from the naive coordinate displacements
$\Delta\varphi$ and $\Delta\chi$, being suppressed when $\lambda_1\chi > 0$. For the benchmarks presented above, the total geodesic field-space traversal
is $\mathcal{O}(10)\,M_{\rm Pl}$. This super-Planckian excursion could however be addressed
within a more complete heterotic construction by means of the Kim--Nilles--Peloso (KNP)
alignment mechanism~\cite{Kim:2004rp} and its multi-axion
generalizations~\cite{Choi:2014rja,Long:2014dta}, which achieve an effectively large decay
constant from multiple axions with sub-Planckian fundamental periodicities. Heterotic
compactifications naturally furnish the required ingredients: a multiplicity of axions arising
from the dimensional reduction of the $B$-field and the gauge sector (cf.\ Section~3), together
with the decay constants that control their alignment. We defer a detailed
implementation of this embedding to future work.

{}Future directions include exploring whether additional symmetries or modified couplings could 
stabilize the dilaton sector in WI, as well as investigating alternative dissipative mechanisms 
beyond gauge interactions. It would also be important to extend the present analysis to the full 
perturbation sector, particularly in regimes where both fields contribute dynamically, in order to 
determine the detailed predictions for the scalar and tensor power spectra.

%%%%%%%%%%%%%%%%%%%%%%%%%%%%%%%%%%%%%%%%%
\acknowledgments

The authors thank Stephon Alexander for useful conversations during the beginning of this work.
A.B. was partially funded by STFC. H.B. was supported by the Simons Foundation through
Award No. 896696 during most of the execution of this work and acknowledges support from the Quantum Horizons Alberta initiative.
S.B. was supported in part by the Higgs Fellowship and by the
STFC Consolidated Grant “Particle Physics at the Higgs Centre”. J.C.F. and R.O.R. gratefully acknowledge the hospitality of the Higgs Centre for Theoretical Physics during the development of part of this work.
J.C.F. is funded by the STFC under grant number ST/X001040/1.
R.O.R. is partially supported by research grants from Conselho
Nacional de Desenvolvimento Cient\'{\i}fico e Tecnol\'{o}gico (CNPq), Grant No. 307286/2021-5, and {}Funda\c{c}\~ao Carlos Chagas Filho de Amparo \`a Pesquisa do Estado do Rio de Janeiro (FAPERJ), Grant
No. E-26/200.415/2026. This material is based upon work supported by the U.S. Department of Energy, Office of Science, Office of High Energy Physics of U.S. Department of Energy under grant Contract Number  DE-SC0012567. MWT  acknowledges financial support from the Simons Foundation (Grant Number 929255).

%%%%%%%%%%%%%%%%%%%%%%%%%%%%%%%%%%%%%%%%  
%\section*{References}
\providecommand{\href}[2]{#2}\begingroup\raggedright\endgroup

\end{document}